\documentclass[pra,twocolumn,nofootinbib,
				]{revtex4-1}

\usepackage{graphicx}
\usepackage{url}
\usepackage{amsmath}
\usepackage{amssymb}
\usepackage{bm}
\usepackage{dsfont}
\usepackage{dcolumn}

\newcommand{\rd}{\mathrm{d}}

\newcommand{\Ho}{\hat{H}}
\newcommand{\Uo}{\hat{U}}

\newcommand{\la}{\langle}
\newcommand{\ra}{\rangle}
\newcommand{\lla}{\langle\!\langle}
\newcommand{\rra}{\rangle\!\rangle}

\newcommand{\T}{\textstyle}
\newcommand{\be}{\begin{equation}}
\newcommand{\ee}{\end{equation}}
\newcommand{\bes}{\begin{eqnarray}}
\newcommand{\ees}{\end{eqnarray}}

\begin{document}

\title{Interband heating processes in a periodically driven optical lattice}

\date{July 15, 2016}

\author{Christoph Str\"ater}
\email{cstraeter@pks.mpg.de}
\affiliation{Max-Planck-Institut f\"ur Physik komplexer Systeme, 
			\mbox{N\"othnitzer Str.\ 38, 01187 Dresden, Germany}}
			
\author{Andr\'e Eckardt}
\email{eckardt@pks.mpg.de}
\affiliation{Max-Planck-Institut f\"ur Physik komplexer Systeme, 
			\mbox{N\"othnitzer Str.\ 38, 01187 Dresden, Germany}}

\begin{abstract}
We investigate multi-``photon'' interband excitation processes in an optical lattice that
is driven periodically in time by a modulation of the lattice depth. Assuming the system 
to be prepared in the lowest band, we compute the excitation spectrum numerically. 
Moreover, we estimate the effective coupling parameters for resonant interband excitation 
processes analytically, employing degenerate perturbation theory in Floquet space. 
We find that below a threshold driving strength, interband excitations are suppressed 
exponentially with respect to the inverse driving frequency. For sufficiently low 
frequencies, this leads to a rather sudden onset of interband heating, once the driving
strength reaches the threshold. We argue that this behavior is rather generic and should
also be found in lattice systems that are driven by other forms of periodic forcing. Our
results are relevant for Floquet engineering, where a lattice system is driven
periodically in time in order to endow it with novel properties like the emergence of a
strong artificial magnetic field or a topological band structure. In this context,
interband excitation processes correspond to detrimental heating.
\end{abstract}
%\pacs{}
%\keywords{}

\maketitle

\section{Introduction}
Floquet engineering is a form of quantum engineering, where a system is periodically 
driven in time, such that it behaves as if it was governed by an effective
time-independent Hamiltonian with desired properties. This concept has recently been 
demonstrated successfully in a series of experiments with ultracold atomic quantum gases 
in driven optical lattices \cite{Eckardt16}. This includes the dynamic localization of a
Bose-Einstein condensate in a shaken optical lattice \cite{LignierEtAl07,EckardtEtAl09},
``photon''-assisted tunneling against a potential gradient \cite{SiasEtAl08, IvanovEtAl08, 
AlbertiEtAl09, HallerEtAl10, MaEtAl11} and the dynamic control of the bosonic Mott 
transition in a strongly interacting system \cite{ZenesiniEtAl09}. The concept of Floquet 
engineering becomes particularly relevant, when the driven system acquires properties 
that are qualitatively different from those of the undriven system. A prime example is the 
realization of artificial magnetic fields, where driven charge-neutral atoms behave as if 
they had a charge coupling to an effective magnetic field \cite{StruckEtAl11, 
AidelsburgerEtAl11, StruckEtAl12, StruckEtAl13, AidelsburgerEtAl13, MiyakeEtAl13, 
AtalaEtAl14, JotzuEtAl14, KennedyEtAl15, AidelsburgerEtAl15}. 

The idea of Floquet engineering is based on the fact that the time evolution of a quantum 
system with time-periodic Hamiltonian $\Ho(t)=\Ho(t+T)$ can be expressed in terms of an effective time-independent Hamiltonian \cite{Shirley65, 
Sambe73}. Namely, the unitary time evolution operator over one driving cycle, from time
$t_0$ to time $t_0+T$, can be written like $\exp(-iT\Ho^F_{t_0}/\hbar)$ in terms of a 
hermitian operator $\Ho^F_{t_0}$ often called Floquet Hamiltonian. 
%Namely, the time-evolution operator from time $t_1$ to time $t_2$ can generally 
%be written like $\Uo_F(t_2)\exp\big(-i(t_2-t_1)\Ho_F/\hbar\big)\Uo_F^\dag(t_2)$, where
%$\Uo_F(t)=\Uo_F(t+T)$ is a unitary operator describing a time-periodic ``micromotion''.
However, the very fact that we can formally define an effective time-independent 
Hamiltonian is not enough to make the concept of Floquet engineering work. We also have 
to require that the effective Hamiltonian can be computed theoretically and takes a 
simple form allowing for a clear interpretation. In an extended system of many 
interacting particles this condition will typically not be fulfilled exactly. Roughly 
speaking, the fact that the driving resonantly couples (and, thus, hybridizes) 
energetically distant states makes the effective Hamiltonian an object much more complex 
than a typical time-independent Hamiltonian. As a consequence of this lack of energy 
conservation, it is believed that a generic many-body Floquet system approaches an 
infinite-temperature-like state in the long-time limit \cite{LazaridesEtAl14b,
DAlessioRigol14}. Floquet engineering, nevertheless, works in an approximate sense in 
parameter regimes, where unwanted resonant coupling is weak and can be neglected on the 
time scale of the experiment.  

In the optical lattice experiments mentioned above, this parameter regime is 
characterized by two conditions \cite{EckardtEtAl05b}. The first one is a
\emph{low-frequency condition}: In order to describe the system in terms of a
Hubbard-type tight-binding model with a single Wannier-like orbital in each lattice 
minimum, one requires the driving frequency to be small compared to the energy gap
$\Delta$ that separates excited orbital degrees of freedom,\footnote{If the lattice 
possesses several minima per elementary cell, like in a hexagonal lattice, the low-energy 
tight-binding model describes a group of several Bloch bands, which is separated by a 
large energy gap $\sim\Delta$ from neglected bands originating from excited on-site orbital degrees of freedom.}
\be\label{eq:low}
\hbar\omega\ll\Delta .
\ee
The second requirement is a \emph{high-frequency condition}: In order to suppress 
resonant coupling within the subspace of low-energy orbitals described by the
tight-binding model, the driving frequency shall be large compared to the matrix element
$J$ for tunneling between neighboring lattice minima and the Hubbard parameter $U$ 
describing on-site interactions,\footnote{Apart from the off-resonance conditions
(\ref{eq:low}) and (\ref{eq:high}), one might also require resonance conditions for 
selected processes. For example, ``photon''-assisted tunneling can be achieved by 
requiring the energy off-sets between neighboring lattice sites to be given by
$\hbar\omega$ \cite{EckardtHolthaus07}.}
\be\label{eq:high}
\hbar\omega \gg U,J. 
\ee
If resonant coupling both to excited orbital states and within the low-energy
tight-binding subspace can be neglected, one can compute the approximate effective 
Hamiltonian relevant for Floquet engineering from the driven tight-binding model using a 
high-frequency expansion \cite{GoldmanDalibard14, BukovEtAl15, EckardtAnisimovas15, 
MikamiEtAl16}. This is the standard approach of Floquet engineering, on which 
the above mentioned optical-lattice experiments are based. 

However, both conditions (\ref{eq:low}) and (\ref{eq:high}) do not completely prevent 
unwanted resonant excitation processes, which in the context of Floquet engineering must 
be viewed as heating. It is therefore crucial to identify the most dominant of these 
heating processes and to estimate their rates. The validity of the high-frequency 
approximation, neglecting resonant excitations within the low-energy Hubbard description, 
has been studied for various scenarios in references \cite{EckardtEtAl05b,
EckardtHolthaus07, EckardtHolthaus08b, PolettiKollath11, EckardtAnisimovas15,
GenskeRosch15, BilitewskiCooper15, BilitewskiCooper15b, BukovEtAl15, CanoviEtAl16}. 
For systems with local energy bound, which includes the fermionic Hubbard model, it has 
been shown that the heating rates decrease exponentially with the driving frequency 
\cite{KuwaharaEtAl16, MoriEtAl16, AbaninEtAl16, AbaninEtAl16b}.
In this paper, we will address the validity of the low-frequency approximation, where the 
resonant coupling to excited orbital states is neglected. Previous work includes 
theoretical studies of resonant inter-orbital coupling due to both single-particle 
processes \cite{DreseHolthaus97, LackiZakrezewski13, Holthaus15} and two-particle scattering
\cite{Sowinski12, ChoudhuryMueller14, ChoudhuryMueller15}. Recently multi-photon interband 
excitations have also been observed experimentally and explained theoretically by
single-particle processes \cite{WeinbergEtAl15}. 

In the following we will systematically investigate heating due to single-particle
multi-photon interband excitation processes in a one-dimensional optical lattice that is 
driven by a modulation of the lattice depth like in the experiments described in references 
\cite{AlbertiEtAl09, MaEtAl11}. For that purpose, we will study interband excitation processes 
numerically and compare these results to analytical estimates that we obtain using perturbation 
theory within the Floquet picture. The latter indicate that heating rates are suppressed 
exponentially for small driving frequencies as long as the driving amplitude remains below a 
threshold value. 

\section{System}

\begin{figure}[t]
\includegraphics[width=1\linewidth]{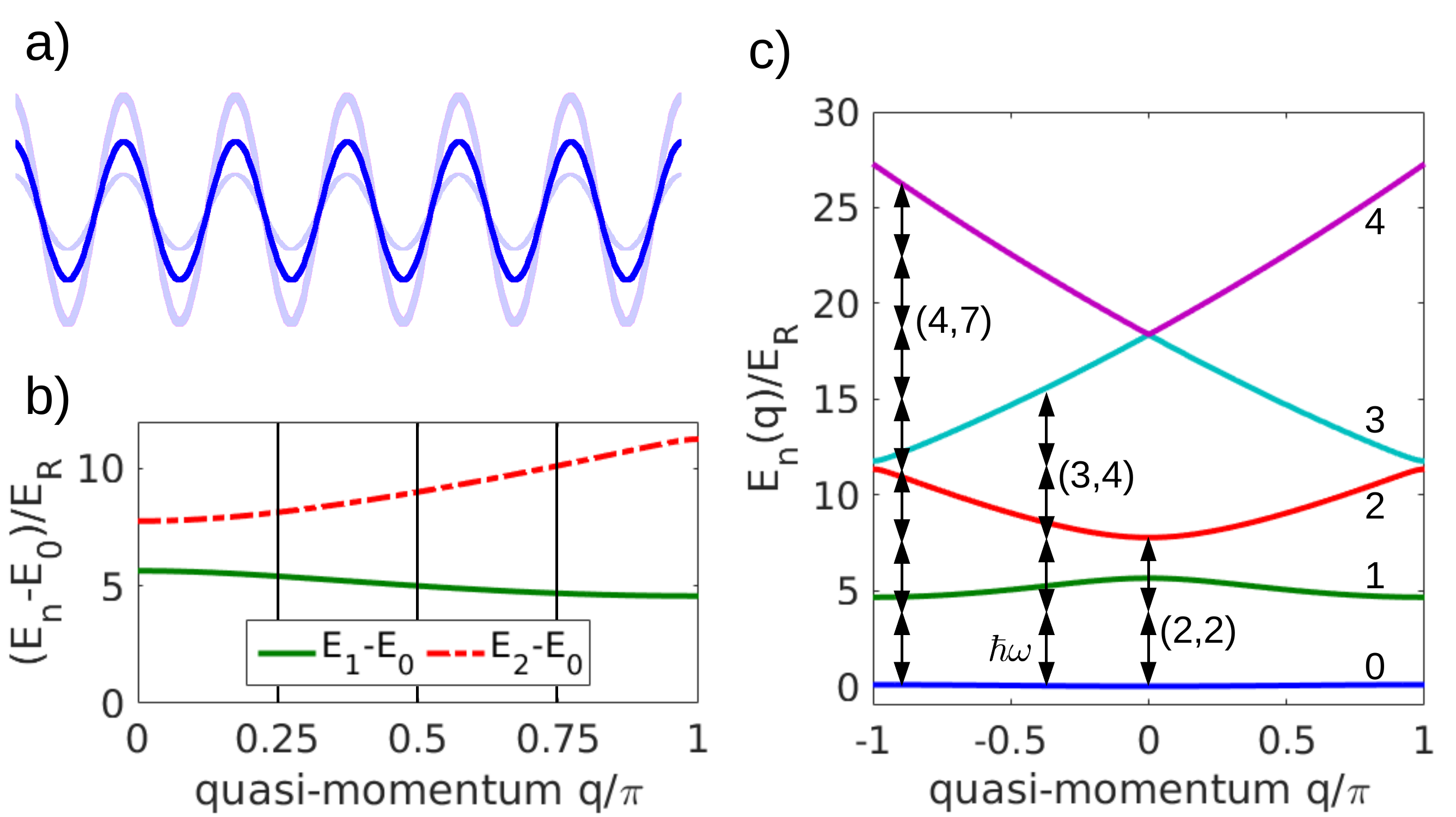}
\centering
\caption{\label{fig:system} Optical lattice of modulated lattice depth (a). Energy 
difference between the first/second excited band and the lowest band versus
quasimomentum (b) and band structure (c) for a static lattice of depth $V_0/E_R=10$.
The label $(b,n)$ denotes the resonance condition $E_n(k)-E_0(k)=n\hbar\omega$ for an $n$
``photon'' transition from the lowest to the $b$th band. } 
\end{figure}

We consider ultracold atoms in a one-dimensional optical lattice, with the lattice depth 
modulated sinusoidally in time [Fig.~\ref{fig:system}(a)]. It is described by the
single-particle Hamiltonian 
\be\label{eq:H}
\Ho(t) = -\frac{\hbar^2}{2m}\partial_x^2 +V_0[1+\alpha\sin(\omega t)]\sin^2(k_Lx)
\ee
where $m$ denotes the particle mass, $V_0$ the average lattice depth, $\alpha$ the 
dimensionless amplitude of the modulation of the lattice depth. The lattice constant
$a=\pi/k_L$ is determined by the wave number $k_L$ of the laser used to create the optical 
lattice. Using the recoil energy $E_R=\frac{\hbar^2k_L^2}{2m}$, corresponding to the 
kinetic energy needed to localize a particle on the length $a$, as the unit of energy the 
system is described by three dimensionless parameters, the lattice depth
$V_0/E_R$, the driving amplitude $\alpha$ and the driving frequency
$\hbar\omega/E_R$. For convenience, we assume periodic boundary conditions, with $M$ 
denoting the number of lattice sites. Since we are interested in single-particle 
excitation effects, we do not need to specify the potential along spatial directions 
other than $x$ by assuming the transverse dynamics separates.

The invariance of the lattice potential with respect to discrete translations $x\to x+a$ 
implies that quasimomentum $q$, i.e.\ momentum $p$ modulo the reciprocal lattice constant 
$2\pi/a$, is conserved. Thus, when describing the system in the basis of momentum 
eigenstates $|p\ra$ with wave functions
\be
\la x|p\ra = \frac{1}{\sqrt{Ma}} e^{ipx}, 
\ee
it is convenient to decompose the momentum wave number like
\be
p = q + \beta \frac{2\pi}{a}, \quad \text{with}\quad -\frac{\pi}{a}<q\le\frac{\pi}{a}
\quad\text{and}\quad \beta\in\mathbb{Z}.
\ee
The wavenumber $q$ can take discrete values that comply with the boundary conditions of the 
system. With respect to the momentum eigenstates, the Hamiltonian possesses the matrix elements
\be
\big\la q'+\beta'\frac{2\pi}{a}\big|\hat{H}'(t)\big
	|q+\beta \frac{2\pi}{a} \big\ra
	= \delta_{q',q}\,  h_{\beta'\beta}(q,t)E_\text{R},
\ee
that are diagonal with respect to $q$, where
\bes\label{eq:matrix}
h_{\beta'\beta}(q,t) &=& \delta_{\beta'\beta}(qa/\pi + 2\beta)^2
\\\nonumber&& +\,\frac{1}{4}\frac{V_0}{E_R}[1+\alpha\sin(\omega t)]
			(\delta_{\beta',\beta+1}+\delta_{\beta',\beta-1}).
\ees

The eigenstates 
\be
|bq\ra=\sum_\beta u_{b\beta}(q)|q+\beta 2\pi/a\ra
\ee
of the undriven Hamiltonian ($\alpha=0$) are labeled by the quasimomentum quantum number
$q$ and the band index $b=0,1,2,\ldots$. Their coefficients $u_{b\beta}(q)$ and energies
$E_b(q)$ are deterimined by the eigenvalue problem
\be
\sum_{\beta'} E_R h_{\beta\beta'}(q) u_{b\beta'}= E_b(q) u_{b\beta}.
\ee
Their wave functions are Bloch waves given by
$
\la x|bq\ra = e^{iqx}\sum_\beta u_{b\beta}(q) e^{i\beta (2\pi/a) x}
			\equiv e^{iqx}u_{bq}(x) 
$,
with $u_{bq}(x+a)=u_{bq}(x)$. The band structure $E_b(q)$ of the undriven system with
$V_0/E_R=10$ is plotted in Fig.~\ref{fig:system}(c). Figure \ref{fig:system}~(b), 
moreover, shows the energy differences between the lowest band the first two excited bands.

\section{Excitation spectrum}

We now assume that the system is initially prepared in a Bloch state $|0q\ra$ of the 
lowest band and investigate excitations to higher-lying bands when the 
driving is switched on at $t=0$. For that purpose we integrate the time-dependent 
Schr\"odinger equation
\be
i\hbar \dot u_\beta(t) = E_R \sum_{\beta'} h_{\beta'\beta}(q,t) u_{\beta'}(t)
\ee
over a time span of $\Delta t$, starting from the initial state
$u_\beta(t=0)=u_{0\beta}(q)$. During the time evolution the state of the system is
given by
\be
|\psi(t)\ra = \sum_\beta u_\beta(t) \big|q+\beta \frac{2\pi}{a} \big\ra.
\ee
Assuming the recoil energy of $E_R=3.33\cdot2\pi\hbar $ kHz, which is a typical value for 
experiments with Rubidium 87 atoms, we choose a time span $\Delta t = 20$ ms.

\begin{figure*}[t]
\includegraphics[width=0.58\linewidth]{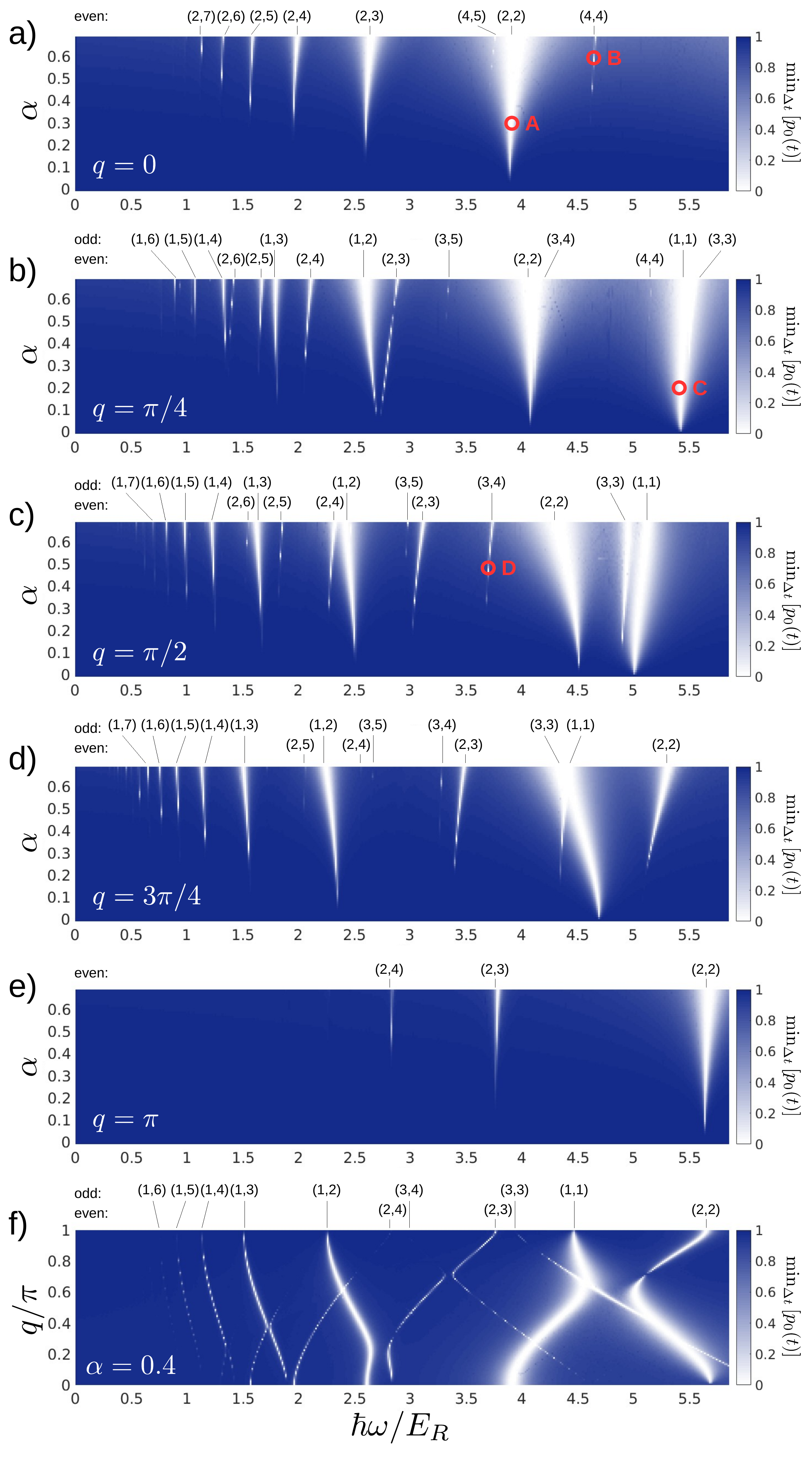}
\centering
\caption{\label{fig:spectrum} Excitation spectrum: Minimum probability
$\min_{\Delta t}[p_0(t)]$ to occupy the initial state $|0q\ra$ during a time span of
$\Delta t = 20$ ms, plotted versus driving frequency $\hbar\omega/E_R$ and either driving 
amplitude $\alpha$ (a-e) or quasimomentum $q$ (f). The parameters are
$V_0/E_R=10$, $E_R=3.33\cdot2\pi\hbar$ kHz, and  $q$ or $\alpha$ as indicated in each
panel. Resonances corresponding to an $n$-``photon'' transition from band 0 to 
$b$ are visible as white stripes and labeled by $(b,n)$. For the points marked 
by $A$, $B$, $C$, $D$, the evolution of the probabilities $p_b(t)$ is depicted in
panels (a), (b), (c), (d) of Fig.~\ref{fig:evolution}, respectively.} 
\end{figure*}

\begin{figure}[t]
\includegraphics[width=0.8\linewidth]{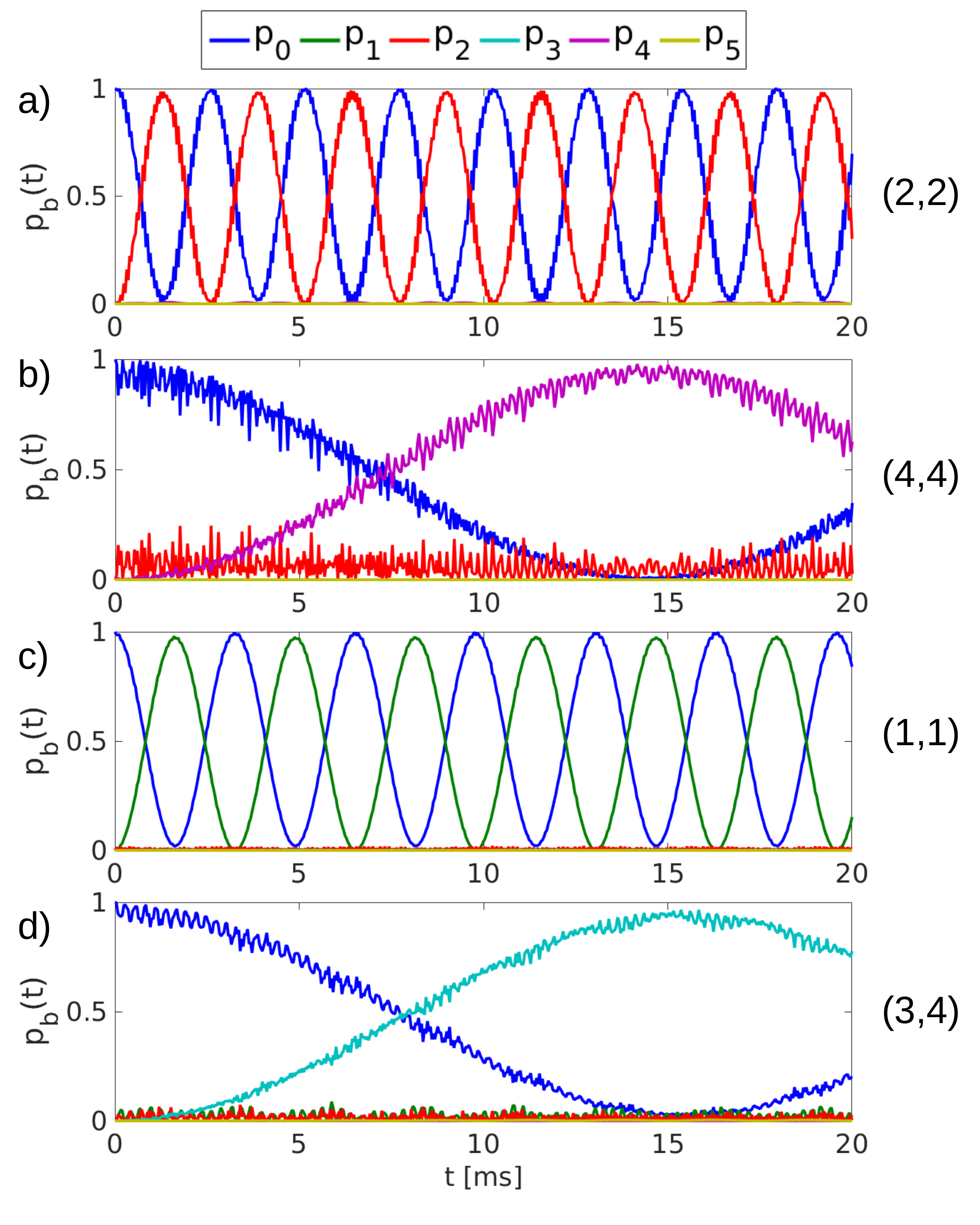}
\centering
\caption{\label{fig:evolution}Time evolution of the populations $p_b(t)$ of the six lowest 
bands. From top to bottom the four subplots correspond to the parameters marked by A, B, 
C, and D marked in Fig.~\ref{fig:spectrum}, respectively.} 
\end{figure}

The probability to find the system in band $b$ is given by the squared overlap 
\be
p_b(t)=|\la \psi(t)|bq\ra|^2=\Big|\sum_\beta u^*_\beta(t) u_{b\beta}(q)\Big|^2.
\ee
In Fig.~\ref{fig:spectrum} we plot the minimal overlap $\min_{\Delta t}[p_0(t)]$ with 
lowest band recorded during the time span $\Delta t$ versus the driving frequency
$\hbar\omega/E_R$ and either the driving amplitude $\alpha$  or the quasimomentum $q$ for
$V_0/E_R=10$. We can clearly observe resonances, where a light color indicates a 
significant transfer out of the lowest band. We have labeled $n$-``photon'' resonances to
band $b$ by $(b,n)$. Such a resonance is expected to occur, roughly, when
\be\label{eq:res}
n\hbar\omega \approx E_n(q)-E_0(q).
\ee 
This resonance condition is also illustrated in Fig.~\ref{fig:system}(c). The precise 
position of the resonance shifts, however, with increasing driving strength, 
since the band structure is effectively modified (dressed) by the periodic forcing.

The character of a resonance can not only inferred from the frequency where it occurs by  
using the resonance condition (\ref{eq:res}). It can also be identified from the time 
evolution of the probabilities $p_b(t)$ for occupying band $b$. In
Fig.~\ref{fig:evolution}, we plot $p_b(t)$ of the six lowest bands versus time. From 
top to bottom panels (a) to (d) of the figure are obtained for the parameters marked by
A, B, C, and D in Fig.~\ref{fig:spectrum}, corresponding to the resonances
$(2,3)$, $(4,3)$, $(1,1)$, and $(3,4)$, respectively. One can clearly identify population
transfer to the bands $b=2$, 4, 1, 3, respectively, as expected from the resonance 
condition (\ref{eq:res}) for $(b,n)$ transitions. From the period $T_{(b,n)}(q)$ of the
oscillations found at a particular resonance, we can define an effective coupling parameter
\be\label{eq:C}
C_{(b,n)}(q)=\frac{2\pi\hbar}{T_{(b,n)}(q)}.
\ee 
When plotting the quasienergy spectrum of the driven system, the resonant coupling between 
different Bloch bands is reflected by the appearance of avoided level crossings. The width of 
the avoided crossing corresponding to the resonance $(b,n)$ is of the order of the coupling 
parameter $C_{(b,n)}$.

The fact that we can see almost full coherent population transfer in the time evolution 
shown in Fig.~\ref{fig:spectrum} is a consequence of the fact that we have chosen the 
parameters to lie precisely where an isolated resonance occurs. When tuning the frequency 
away from the resonance, so that the detuning becomes comparable to the effective 
coupling parameter $C_{(b,n)}(q)$, oscillations with incomplete population transfer 
occur. When the detuning becomes much larger than $C_{(b,n)}(q)$, significant population 
transfer is suppressed. Thus, in the spectra of Fig.~\ref{fig:spectrum}, the width of a 
resonance feature reflects the effective coupling matrix element $C_{(b,n)}(q)$ related to the 
excitation process.  Additionally, for small effective coupling matrix elements
$C_{(b,n)}(q)< \pi \hbar/\Delta t$ the oscillations 
are truncated by the finite integration time $\Delta t$. In the spectra of
Fig.~\ref{fig:spectrum}, this effect leads to resonances dips with the minimum taking 
values larger than zero. Thus, resonances with $C_{(b,n)}(q)\ll \pi \hbar/\Delta t$, 
which are not relevant on the time scale $\Delta t$ are suppressed.

From the excitation spectra shown in Fig.~\ref{fig:spectrum}, we can infer some general 
trends. \emph{(i)} The resonances tend to become broader with increasing driving strength
$\alpha$.\footnote{Apparent oscillations, as they are visible in thin resonance lines 
like $(2,3)$ in panel (b) of Fig.~\ref{fig:spectrum}, are an artifact of the finite 
frequency resolution of the underlying data.} This observation is not surprising, since 
the resonant coupling is induced by the driving. \emph{(ii)} The lower the frequency, i.e.\
the larger the number $n$ of ``photons'' required, the weaker is the resonant coupling to 
a given band $b$. (iii) In the limit of low driving frequencies, the resonance features 
disappear abruptly, when the driving strength $\alpha$ falls below a finite threshold, 
which decreases for increasing driving frequency. \emph{(iv)} Resonances to bands with
odd index $b$ are completely suppressed for the quasimomenta $q=0$ and $q=\pi/a$. For 
other quasimomenta, they exist, but they are systematically weaker than even resonances.
This can be seen for example in panel (c) of Fig.~\ref{fig:spectrum}. Here the $n$-photon
transitions to the first band $(1,n)$ give rise to a narrower dips than corresponding
transition to the second band with the same $n$, $(2,n)$. \emph{(v)} Within the groups
of even and odd bands, for a given ``photon'' number $n$ resonances to higher excited
bands tend to be weaker than resonances to lower bands. 
For example in Fig.~\ref{fig:spectrum}(a) the $(4,4)$ resonance is much weaker than the
$(2,4)$ resonance and in Fig.~\ref{fig:spectrum}(d) the $(3,5)$ resonance is much weaker
than the $(3,1)$ resonance. Transitions to higher-lying bands are suppressed, 
furthermore, by the larger excitation energy requiring a larger number $n$ of ``photons'' for a 
given frequency $\hbar\omega/E_R$. In order to justify the low-frequency approximation it is, 
therefore, most crucial to study transitions to the lowest even and odd band, $b=2$ and $b=1$, 
since these give rise to the strongest resonances for a given frequency regime. In the 
following section we will estimate the effective coupling matrix elements $C_{(b,n)}(q)$ 
using analytical arguments. This will allow us to explain the observations made on the 
basis of the numerically computed excitation spectra.

\section{Estimating the effective coupling parameter}

\subsection{Hamiltonian}
As a prerequisite for further investigation, it is convenient to perform a gauge 
transformation
\bes
|\psi(t)\ra &\to& |\psi'(t)\ra=\Uo^\dag(t)|\psi(t)\ra 
\\
\Ho(t)&\to&\Ho'(t)=\Uo^\dag(t)\Ho(t)\Uo(t)-i\hbar\Uo^\dag(t)\dot\Uo(t)
\ees
with the time-periodic unitary operator
\be\label{eq:trans}
\Uo(t)=\sum_{q,b}|bq,t\ra\la bq| .
\ee
Here we have introduced the normalized instantaneous eigenstates $|bq,t\ra$ of the
time-dependent Hamiltonian,
\be
\Ho(t)|bq,t\ra = E_b(q,t)|bq,t\ra.
\ee
They are Bloch waves of the lattice system at the instantaneous lattice depth
$V_0[1+\alpha\sin(\omega t)]$ labeled by the same quantum numbers, quasimomentum $q$ and 
band index $b$, as the eigenstates of the undriven system. 
The transformed Hamiltonian reads
\be
\Ho'(t) = \sum_q\Ho'(q,t)
\ee
with
\be
\Ho'(q,t) = \sum_b |bq\ra E_b(q,t)\la bq| 
		+\sum_{bb'} |b'q\ra M_{b'b}(q,t)\la bq| 
\ee
and matrix elements
\be
M_{b'b}(q,t) = -i\hbar\la b'q,t|\partial_t|bq,t\ra. 
\ee

For the sake of a light notation, in the following we will suppress the quasimomentum 
label $q$, when denoting states, energies, and matrix elements. Applying the 
transformation (\ref{eq:trans}) is a standard procedure when treating slow parameter 
variations in quantum systems. Following this standard procedure further, we can bring 
the matrix elements $M_{b'b}(t)$ in a more convenient form. Let us first 
discuss the diagonal matrix elements. They describe Berry phase effects and can, in the 
present case, be removed by a simple gauged transformation, since we are varying a single 
parameter, the lattice depth, during each driving cycle only.  
Namely, we can write the diagonal matrix elements like
\be
M_{bb}(t) = -i\hbar\la b,t|\partial_t|b,t\ra = -\hbar A_b(V)\dot V(t)    
\ee
in terms of the Berry connection $A_b(t) = i\la b,V|\partial_V|b,V\ra$ for a variation of 
the lattice depth $V$. Here we have introduced the eigenstates $|b,V\ra$ for a lattice of 
depth $V$, so that $|b,t\ra=|b,V(t)\ra$ with $V(t)=V_0[1+\alpha\sin(\omega t)]$.
A gauge transformation $|b,V\ra'=e^{i\theta_b(V)}|b,V\ra$ changes the Berry curvature to
$A_b'(V)=A_b(V)-\partial_V\theta_b(V)$, which vanishes for the choice
$\theta_b(V)=\int_0^V\!\rd W\, A_b(W)$. Thus, for a suitable definition of the phase of the 
instantaneous eigenstates, the diagonal matrix elements vanish
\be
M_{bb}(t) = 0.
\ee
Berry phase effects can matter, however, in more complicated driving scenarios involving 
the variation of several parameters. 

In order to evaluate the off diagonal matrix elements $M_{b'b}(t)$ with $b'\ne b$, we 
consider the quantity $\la b',t|\frac{\rd}{\rd t}(\Ho'(t)|b,t\ra)$, which can be evaluated 
to both $\la b',t|\dot \Ho(t)|b,t\ra+E_{b'}(t)\la b',t|\partial_t|b,t\ra$ and 
$E_b(t)\la b',t|\partial_t|b,t\ra$. Equating both provides an expression for
$\la b',t|\partial_t|b,t\ra$ that gives 
\be\label{eq:M}
M_{b'b}(q,t) = -i\frac{\hbar\la b'q,t|\dot\Ho(t)|bq,t\ra}{E_{b'}(q,t)-E_b(q,t)}
\ee
as long as $E_{b'}(q,t)\ne E_b(q,t)$. Here we have reintroduced the quasimomentum $q$. 

All in all, the system is described by the time-periodic Hamiltonian 
\be
\Ho'(q,t) =\sum_b \Big[ |bq\ra E_b(q,t)\la bq| 
		+\sum_{b'\ne b} |b'q\ra M_{b'b}(q,t)\la bq| \Big].
\ee
So far, no approximation has been made. 

\begin{figure}[t]
\includegraphics[width=0.8\linewidth]{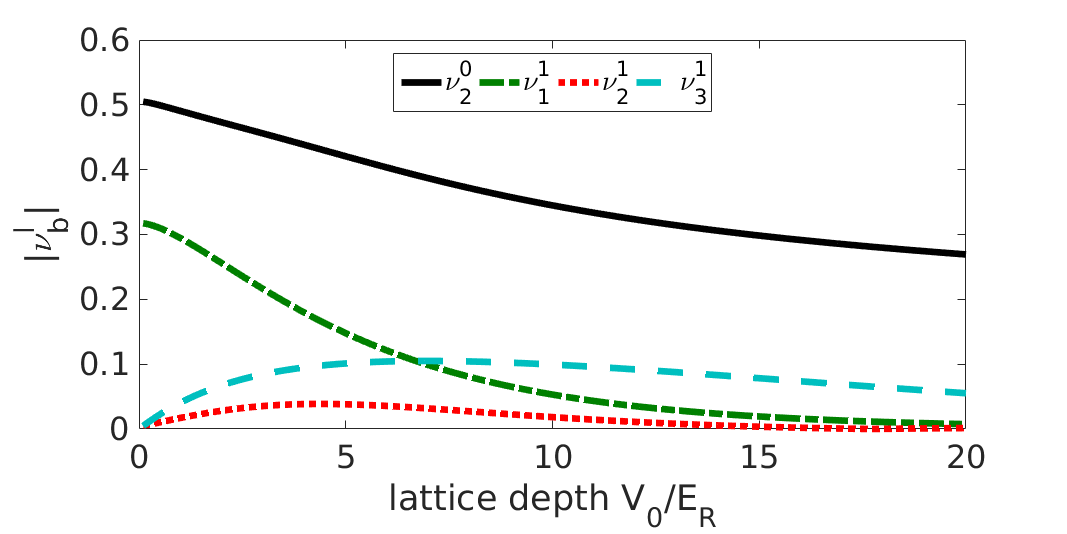}
\centering
\caption{\label{fig:coupling}Coupling matrix elements $\nu^\ell_{b}=W_{b0}^{(\ell)}$ as 
defined in Eq.~(\ref{eq:MatrixElements}) for a static lattice of depth $V_0/E_R$.} 
\end{figure}

The properties of the matrix elements (\ref{eq:M}) become more transparent, when 
expressing the instantaneous Bloch waves in terms of instantaneous Wannier
states~$|b\ell,t\ra$,
\be
|bq,t\ra = \frac{1}{\sqrt{M}}\sum_\ell e^{iqa\ell} |b\ell,t\ra.
\ee
Their wave functions 
\be
\la x|b\ell,t\ra = w_b(x-\ell a,t) 
\ee
are real and exponentially localized at the lattice minima $x=\ell a$ with integer
$\ell$; moreover, $w_b(x)$ is even (odd) for $b$ even (odd), $w_b(-x)=(-1)^b w_b(x)$ 
\cite{Kohn59}. The time dependence describes a breathing motion of the Wannier 
functions, since the width of the Wannier orbitals decreases slightly with increasing 
lattice depth. The numerator on the right-hand side of Eq.~(\ref{eq:M}) can then be 
expressed like 
\be\label{eq:numerator}
\hbar\la b'q,t|\dot\Ho(t)|bq,t\ra 
	= \alpha V_0\hbar\omega \cos(\omega t) \sum_\ell e^{iqa\ell} W_{b'b}^{(\ell)}(t),
\ee
with matrix elements (see Fig.~\ref{fig:coupling})
\be\label{eq:MatrixElements}
W_{b'b}^{(\ell)}(t)=\int\!\rd x\, w_{b'}(x+\ell a,t) \sin^2(k_L x) w_b(x,t) 
\ee
that obey
\be
W_{b'b}^{(-\ell)}(t)=(-1)^{b+b'} W_{b'b}^{(\ell)}(t).  
\ee
Thus, for even $(b'+b)$ the sum on the right-hand side of Eq.~(\ref{eq:numerator}) reads
\be\label{eq:Weven}
W_{b'b}^{(0)}(t)+2 W_{b'b}^{(1)}(t)\cos(qa)+2W_{b'b}^{(2)}(t)\cos(2qa)+\cdots,
\ee
whereas for odd $(b'+b)$ the leading $\ell=0$ term vanishes and one finds
\be\label{eq:Wodd}
2i W_{b'b}^{(1)}(t)\sin(qa)+2iW_{b'b}^{(2)}(t)\sin(2qa)+\cdots . 
\ee
These equations explain why transitions to odd bands are suppressed completely in the 
spectra of Fig.~\ref{fig:spectrum} for $q=0$ and $q=\pi/a$. The missing $\ell=0$ term for 
odd transitions, which is related to parity conservation within a single lattice site, 
explains also the observed relative suppression of transitions from the lowest to odd 
bands for other values of $q$. Namely, due to the exponential localization of the Wannier 
functions, the matrix elements $W_{b'b}^{(\ell)}(t)$ drop rapidly with $\ell$. It is, 
therefore, reasonable to keep only the leading term and to approximate
\be
\hbar\la b'q,t|\dot\Ho(t)|bq,t\ra 
	= \alpha V_0\hbar\omega \cos(\omega t) W_{b'b}^{(0)}(t)
\ee
for even $(b'+b)$ and
\be
\hbar\la b'q,t|\dot\Ho(t)|bq,t\ra 
	= i 2\alpha V_0\hbar\omega \cos(\omega t)W_{b'b}^{(1)}(t)\sin(qa)
\ee
for odd $(b'+b)$.

In order to be explicit, in the following we will focus on transitions from the lowest to 
the second excited band. For small quasimomenta $k\ll\pi/a$ these transitions constitute 
the dominant heating channel. The relevant matrix element $W_{20}^{(0)}$ has a rather 
weak dependence on the lattice depth, as can be seen from Fig.~\ref{fig:coupling} where this 
parameter is plotted for a lattice of static depth $V_0$. Thus, when the lattice depth is 
modulated, $V_0\to V_0[1+\alpha\sin(\omega t)]$, we can approximate
\be
W_{20}^{(0)}(t)\approx W - \alpha W' \sin(\omega t),
\ee
neglecting higher harmonics. Both coefficients $W$ and $W'$ have a very weak 
dependence on $\alpha$ only and one has $W'\ll W\sim 1$. At an $n$``photon'' resonance, 
we can likewise approximate the instantaneous energy difference between both bands like 
\be
E_2(q,t)-E_0(q,t) \approx \Delta(q)  + \alpha F(q)\sin(\omega t)  
\ee
and its inverse like
\be
\frac{1}{E_2(q,t)-E_0(q,t)} \approx \frac{1}{\Delta(q)}
					- \alpha \frac{F(q)}{\Delta^2(q)}\sin(\omega t) .
\ee
Taking terms up to $\alpha^2$, the matrix element $M_{20}(q,t)$ then reads
\be
M_{20}(q,t)\approx -i \frac{V_0}{n}\big[\alpha W \cos(\omega t) 
	 -\alpha^2 X(q)	\sin(2\omega t)\big],
\ee
where we have used $\sin(a)\cos(a)=\sin(2a)/2$, employed the resonance condition
\be\label{eq:resonance}
\Delta(q) = n\hbar\omega,
\ee
and defined
\be\label{eq:X}
X(q)=\frac{1}{2}\bigg[W' +\frac{F(q)}{\Delta(q)}\bigg],
\ee
where $X(q)\ll W$. 

At the resonance $(2,n)$, we will describe the system within the the subspace spanned by the bands $b=0$ and $b=2$. Up to 
a time-dependent energy constant, the relevant Hamiltonian is given by 
\bes\label{eq:Hp}
\Ho'(q,t) &\approx& \big[n\hbar\omega + \alpha F(q)\sin(\omega t)\big] |2q\ra\la 2q|
\\\nonumber&&
		+\, M_{20}(q,t) |2q\ra\la 0q| + M_{20}^*(q,t) |0q\ra\la 2q|  .
\ees
The Fourier decomposition of the Hamiltonian is given by
\bes
\Ho'(q,t) &=& \sum_m \Ho'_m(q) e^{im\omega t} ,
\\
\Ho'_m(q)&=&\frac{1}{T}\int_0^T\!\rd t\,e^{-im\omega t} \Ho'(q,t),
\ees
with driving period $T=2\pi/\omega$, we find 
\bes\label{eq:H0}
\Ho'_0(q) &=& n\hbar\omega |2q\ra\la 2q|,
\\\label{eq:H1}
\Ho'_1(q) &=& -i\frac{\alpha F(q)}{2} |2q\ra\la 2q|
\\\nonumber\\&& 
		 -i \,\frac{\alpha V_0W}{2n} \big(|2q\ra\la 0q| - |0q\ra\la 2q|\big),
\\\label{eq:H2}
\Ho'_2(q) &=& \frac{\alpha^2 V_0X(q)}{2n}
						\big(|2q\ra\la 0q| - |0q\ra\la 2q|\big),
\ees
as well as the conjugated terms $\Ho_{-m}=\Ho_m^\dag$. The terms $\Ho_m$ become smaller 
with increasing $m$ and depend on the driving strength like $\alpha ^{|m|}$. This applies 
also to the higher harmonics that we neglected.

\subsection{Rotating-wave approximation}
If the coupling matrix element $M_{20}(q,t)$ is small compared to the driving frequency
$\hbar\omega$ (both scale like $1/n$), a rotating wave approximation is justified. For 
this approximation, we perform yet another gauge transformation, with the unitary operator
\be
\Uo'(t)=\exp\Big(-i\sum_q\Big[ n\omega t 
		-\frac{\alpha F(q)}{\hbar\omega}\cos(\omega t)\Big]|2q\ra\la 2q|\Big).
\ee
Assuming the resonance condition $(\ref{eq:resonance})$, the transformed Hamiltonian reads
\be\label{eq:Hpp}
\Ho''(q,t) = M_{20}(q,t)e^{in\omega t 
		- i\frac{\alpha F(q)}{\hbar\omega}\cos(\omega t)}|2q\ra\la 0q|
		+\text{ h.c. }.
\ee
In the following, we will again drop the label $q$. Employing the relation
\be
\exp(-ia\cos(b)) = \sum_{k=-\infty}^\infty(-i)^k \mathcal{J}_k(a)  e^{-ikb},
\ee
where $\mathcal{J}_k(x)$ denotes a Bessel function of the first kind, we find the Fourier 
components of the time-dependent matrix element
\be
M_{20}(q,t)e^{in\omega t 
		- i\frac{\alpha F(q)}{\hbar\omega}\cos(\omega t)}
		=\sum_r M^{(n)}_r e^{ir\omega t}
\ee
to be given by
\bes\label{eq:Mr}
M^{(n)}_r &=& -i\frac{\alpha V_0 W}{2n}
	(-i)^{n+1-r}\mathcal{J}_{n+1-r}\big({\T \frac{\alpha F}{\hbar\omega}}\big)
\\\nonumber&&\quad
	\,-i\frac{\alpha V_0 W}{2n}
	(-i)^{n-1-r}\mathcal{J}_{n-1-r}\big({\T \frac{\alpha F}{\hbar\omega}}\big)
\\\nonumber && 
		+\,\frac{\alpha^2 V_0 X}{2n}
	(-i)^{n+2-r}\mathcal{J}_{n+2-r}\big({\T \frac{\alpha F}{\hbar\omega}}\big)
\\\nonumber&&
	-\,\frac{\alpha^2 V_0 X}{2n}
	(-i)^{n-2-r}\mathcal{J}_{n-2-r}\big({\T \frac{\alpha F}{\hbar\omega}}\big).
\ees
For the rotating-wave approximation, we now neglect the rapidly rotating phases of the
coupling matrix element,
\be
M_{20}(q,t)e^{in\omega t 
		- i\frac{\alpha F(q)}{\hbar\omega}\cos(\omega t)}\approx M^{(n)}_0.
\ee
The effective coupling parameter is, thus, given by 
\be\label{eq:Crw}
C_{(2,n)} =|M^{(n)}_0|.
\ee

In order to interpret this result, it is useful to make further approximations. 
First of all, let us consider only the leading order with respect to the driving strength 
$\alpha$. For this purpose, we note that for small arguments $x$ (and $k\ge0$)
the Bessel function is asymptotically given by 
\be
\mathcal{J}_k(x)\simeq \frac{1}{k!}\Big(\frac{x}{2}\Big)^k. 
\ee
Hence, in leading order only the second and the fourth line of Eq.~(\ref{eq:Mr}) 
contribute to $M_0$ and we have
\be\label{eq:Mn}
M^{(n)}_0 \simeq (-i)^n \frac{\alpha V_0}{n}
\bigg[\frac{W}{2}+\frac{(n-1)X\hbar\omega}{F}\bigg]
\frac{\Big(\frac{\alpha F}{2\hbar\omega}\Big)^{\!n-1}}{(n-1)!}
\ee
For large ``photon'' numbers $n$, we can now use Stirling's formula 
\be\label{eq:Stirling}
k! \simeq \sqrt{2\pi k}\Big(\frac{k}{e}\Big)^k
\ee
valid for large $k$. We obtain
\be\label{eq:M0}
C_{(2,n)}\simeq 
\alpha V_0 \sqrt{\frac{\pi}{2n^3}}\bigg(W
			+\frac{W'\Delta}{F}+1\bigg)
\bigg(\frac{\alpha }{\alpha_\text{thresh}}\bigg)^{\!n-1},
\ee
with the threshold value
\be\label{eq:thresh}
\alpha_\text{thresh} = \frac{2\Delta}{eF}
\ee
for the driving strength. Here we also employed Eqs.~(\ref{eq:resonance}) and
(\ref{eq:X}).
We can compare the estimate given by the rotating wave approximation to the numerical computed 
dynamics. 

From the evolution shown in Fig.~\ref{fig:evolution}(a), we can extract the period 
$T^\text{sim}_{(2,2)}\approx 2.56\,\mathrm{ms}$ for $\alpha=0.3$, $\hbar\omega=3.9 E_R$, $q=0$,
$V_0/E_R=10$ and $E_R=3.33\cdot2\pi\hbar \mathrm{kHz}$. For these parameters, we obtain
$\Delta \approx 7.77 E_R$, $F(0)\approx5.51 E_R$, $W \approx 0.345$, as well as $W'\approx 0.12$.
Using Eq.~\ref{eq:C} and the rotating wave approximation for the coupling parameter
(\ref{eq:Mn}) and (\ref{eq:Crw}), we obtain the estimate
$T^\text{RW}_{(2,2)}\approx 2.03 \,\mathrm{ms}$ for the oscillation period, which lies about
twenty percent below the numerically observe value.

Equation (\ref{eq:M0}) tells us that for large $n$ the onset of heating occurs in a 
rather sharp transition when the driving strength reaches the threshold. Namely,
for $\alpha<\alpha_\text{thresh}$ the coupling parameter is exponentially suppressed 
with respect to $n=\Delta/\hbar\omega$. This result is favorable for Floquet engineering, 
as it tells us that for sufficiently low frequencies and not too strong driving, 
interband heating becomes very small. However, the predicted threshold is only valid as
long as $M_{20}(t)$ is small compared to $\hbar\omega$ for $\alpha=\alpha_\text{thresh}$. 
If this is not the case, we have to go beyond the rotating wave approximation. This can 
be done using degenerate perturbation theory in Floquet space.

\subsection{Floquet perturbation theory}
Let us now estimate the effective coupling parameter $C_{(2,n)}(q)$ for the resonant 
$n$-``photon'' coupling of the states $|0q\ra$ and $|2q\ra$ using degenerate perturbation 
theory within the Floquet space of the driven system
(see, e.g., Ref.~\cite{EckardtAnisimovas15}). Within this space the state $|bq\ra$ is
represented by a family of states $|bqm\rra$ labeled by an integer index $m$ that
represent a time-dependent state in the original state space $|bq\ra e^{im\omega t}$. The
coupling between these states, which form a complete basis, is described by the
quasienergy operator $\bar{Q}$ playing the role of a static Hamiltonian. Starting from 
the problem defined by the Hamiltonian $\Ho'(q,t)$, the matrix matrix elements of the quasienergy operator are given by 
\be\label{eq:Q}
\lla b'q'm'|\bar{Q}|bqm\rra = \la b'q'|(\delta_{m'm}m\hbar\omega+\Ho'_{m'-m}(q))|bq\ra,
\ee
where $\Ho'_m(q)$ denote the Fourier components of the Hamiltonian.
The eigenstates and eigenvalues of the quasienergy operator correspond to the
time-periodic Floquet modes and their quasienergies, respectively, which play a role 
similar to that of the stationary states in undriven systems and their energies. 

The integer $m$ plays the role of the relative occupation of a photonic mode in the 
classical limit of large occupation. In this interpretation the state $|bqm\rra$ 
represents a product state $|bq\ra|m\ra$ solving the unperturbed problem 
\bes
\lla bqm'|\bar{Q}_0|bqm\rra &=& \delta_{m'm}\la b'q'|(m\hbar\omega+\Ho_{0}')|bq\ra
\\\nonumber
	&=& \delta_{m'm}\delta_{q'q}\delta_{b'b}\big[m\hbar\omega + \varepsilon_b(q) \big].
\ees
The unperturbed quasienergy $\varepsilon_{bm}(q)$ is thus given by the
``photonic'' energy $m\hbar\omega$ plus the system energy 
\be
\frac{1}{T}\varepsilon_b(q)=\int_0^T\!\rd t\, E_b(q,t),
\ee
$\varepsilon_{bm}(q)=m\hbar\omega+\varepsilon_b(q)$. The coupling between the
``photonic'' mode and the system is described by the Fourier components $\Ho_m$ of the 
Hamiltonian with $m\ne0$,
\bes
\lla bqm'|\bar{V}|bqm\rra&=&\lla bqm'|(\bar{Q}-\bar{Q}_0)|bqm\rra 
\\\nonumber&=& (1-\delta_{m'm})\delta_{q'q}\la b'q|\Ho_{m'-m}(q)|bq\ra.
\ees

In order to describe the coupling between the lowest and the second excited band, let us 
write down the relevant matrix elements of the quasienergy operator explicitly. For the 
sake of a light notation, we will again suppress the label $q$. The diagonal matrix 
elements are given by
\bes
\lla 0m|\bar{Q}_0|0m\rra &=& m\hbar\omega ,
\\
\lla 2m|\bar{Q}_0|2m\rra &=& m\hbar\omega + \varepsilon_2,
\ees
where $\varepsilon_0=0$ was chosen for convenience. At the $n$-``photon'' resonance
$(2,n)$, where the lowest band is resonantly coupled to the second excited band, we have
\be
\varepsilon_2\approx n\hbar\omega,
\ee 
so that the states $|0 m\rra$ and $|2(m-n)\rra$ are (nearly) degenerate. The relevant 
coupling matrix elements of the perturbation $\bar{V}$ change the photon number $m$ by $\pm1$ or by $\pm2$ (so that for $n>2$ necessarily higher-order processes have to be taken into account in order to describe the coupling between $|0 m\rra$ and $|2(m-n)\rra$). They are given by 
\bes\label{eq:V0}
\lla 2(m\pm 1)|\bar{V}|2m\rra &=& \mp i\frac{\alpha F}{2}
\\\label{eq:V1}
\lla 2(m\pm 1)|\bar{V}|0m\rra &=& -i \frac{\alpha V_0W}{2n}
\\\label{eq:V2}
\lla 2(m\pm 2)|\bar{V}|0m\rra &=& \pm \frac{\alpha^2 V_0X}{2n}
\ees
and the hermitian conjugated terms, where we have employed Eqs.~(\ref{eq:H1}) and
(\ref{eq:H2}).

The coupling parameter $C_{(b,n)}$ introduced in Eq.~(\ref{eq:C}) corresponds to the
absolute value of the matrix element coupling the states $|0 m\rra$ and $|2(m-n)\rra$ in 
Floquet space. 
For the single-``photon'' resonance with $n=1$, both states are directly coupled by the matrix element (\ref{eq:V1}), so that the coupling parameter reads 
\be
C_{(2,1)} = |\lla 2(m-1)|\bar{V}|0m\rra|=\frac{\alpha V_0 W}{2}.
\ee

For the two-``photon'' resonance with $n=2$, we have two relevant contributions to the 
coupling parameter,
\be\label{eq:2PT}
C_{(2,2)} = \big|C^{(1)}_{(2,2)} + C^{(2)}_{(2,2)}\big|.
\ee
The first contribution directly corresponds to the matrix element (\ref{eq:V2}) 
describing a two-photon process, 
\be
C^{(1)}_{(2,2)} =\lla 2(m- 2)|\bar{V}|0m\rra= -\frac{\alpha^2 V_0X}{4}.
\ee
The second contribution stems from the second-order processes
$|0m\rra\to|2(m-1)\rra\to|2(m-2)\rra$, where both states are coupled via an energetically 
distant intermediate state. The unperturbed quasienergy of this intermediate state,
$\varepsilon_{2(m-1)}=n\hbar\omega+(m-1)\hbar\omega$, lies $\hbar\omega$ above the quasienergy
$\varepsilon_{0m}=\varepsilon_{2(m-n)}=m\hbar\omega$ of the degenerate doublet. According to 
the rules of degenerate perturbation theory (see, e.g., Ref.~\cite{EckardtAnisimovas15}), we 
find 
\bes
C^{(2)}_{(2,2)} &=& \frac{\lla 2(m-2)|\bar{V}|2(m-1)\rra
				\lla 2(m-1)|\bar{V}|0m\rra}{\varepsilon_{0m}-\varepsilon_{2(m-1)}}
\\\nonumber&=&
	-\frac{\alpha^2FV_0W}{8\hbar\omega}.
\ees

Let us finally have a closer look also at the three ``photon'' process with $n=3$. The 
coupling parameter is a combination of three contributions, 
\be
C_{(2,3)} =\big| C^{(2)}_{(2,3)} + C^{(3a)}_{(2,3)}+ C^{(3b)}_{(2,3)}\big|.
\ee
The first contribution stems from the second-order process
$|0m\rra\to|2(m-2)\rra\to|2(m-3)\rra$. The intermediate state has a quasienergy lying
$\hbar\omega$ above the degenerate doublet and the resulting coupling is given by
\bes
C^{(2)}_{(2,3)} &=& \frac{\lla 2(m-3)|\bar{V}|2(m-2)\rra
				\lla 2(m-2)|\bar{V}|0m\rra}{\varepsilon_{0m}-\varepsilon_{2(m-2)}}
\\\nonumber&=&
	-\frac{\alpha^3FV_0X}{4n\hbar\omega}.
\ees
The second contribution stems from the third-order processes
$|0m\rra\to|2(m-1)\rra\to|0(m-1)\ra\to|2(m-3)\rra$. The quasienergies of both intermediate 
states are separated by $2\hbar\omega$ and $-\hbar\omega$ from the degenerate doublet 
of states to be coupled. The matrix element is, thus, of the order of 
\be
C^{(3a)}_{(2,3)}\sim \bigg(-i\frac{\alpha V_0W}{2n}\bigg)^3
	\frac{1}{(-2\hbar\omega)(\hbar\omega)}
	=-i\frac{\alpha^3V_0^3W^3}{432(\hbar\omega)^2}.
\ee
The third contribution stems, finally, from the third order process 
$|0m\rra\to|2(m-1)\rra\to|2(m-2)\ra\to|2(m-3)\rra$. The quasienergies of both intermediate 
states are separated by $2\hbar\omega$ and $\hbar\omega$ from the degenerate doublet. 
The corresponding coupling parameter is of the order of
\be
C^{(3b)}_{(2,3)}\sim  \bigg(\frac{i\alpha F}{2}\bigg)^2\frac{-i\alpha V_0 W}{2n}
	\frac{1}{(-2\hbar\omega)(-\hbar\omega)}
	=i\frac{\alpha^3F^2V_0W}{32(\hbar\omega)^2}.
\ee

Extending the perturbative arguments used here to higher orders of the perturbation 
theory, one can estimate also the coupling parameters $C_{(2,n)}$ for multi-``photon'' 
transitions with $n>3$. A similar approach can, moreover, also be applied for transitions 
to the first excited band or higher lying bands. In leading order in the driving 
strengths $\alpha$, we can again cast the coupling parameters into the very same form
\be\label{eq:scaling}
C_{(b,n)} = \alpha B_{(b,n)}\bigg(\frac{\alpha}{\alpha_{(b,n)}}\bigg)^{\!n-1} 
\ee
encountered already within the rotating-wave approximation (\ref{eq:M0}), with energy
scale $B_{(b,n)}$ and threshold driving strength $\alpha_{(b,n)}$. This form 
implies that for below-threshold driving, $\alpha<\alpha_{(b,n)}$, interband excitation 
processes are suppressed exponentially for large ``photon'' numbers
$n=\Delta/(\hbar\omega)$, that is for low frequencies. However,
while Eq.~(\ref{eq:scaling}) is of the same form as 
the rotating-wave result, the coefficient $B_{(b,n)}$ and the threshold value
$\alpha_{(b,n)}$ will generally be different.

The characteristic driving strength $\alpha_{(b,n)}$, below which heating is suppressed, 
might show oscillatory behavior between even and odd $n$. 
%Apart from that, we expect the general trend that $\alpha_{(b,n)}$ increases with
%increasing $n$. Such a behavior is clearly visible in the spectra of
%Fig.~\ref{fig:spectrum}, where for lower frequencies significant resonance features are
% visible for stronger driving strengths $\alpha$ only. An increase of $\alpha_{(b,n)}$ 
% with increasing $n$ can also be motivated using arguments based on perturbation theory
Apart from such details, let us estimate how $\alpha_{(b,n)}$ scales when $n$ becomes 
large. For that purpose, the first quantity to be studied are the energy denominators of 
the perturbatively computed coupling parameters. They are given by the product of the
quasienergies that the intermediate states have with respect to the degenerate doublet of
states to be coupled. Taking, for simplicity, a sequence of processes that lower the 
``photon'' number in steps of one, these denominators provide a factor of 
\be
%\frac{1}{(n-1)\hbar\omega\cdot(n-2)\hbar\omega \cdots\hbar\omega}
\frac{1}{(n-1)!(\hbar\omega)^{n-1}} 
\simeq
 \frac{1}{\sqrt{2\pi (n-1)}}\bigg(\frac{e}{(n-1)\hbar\omega}\bigg)^{(n-1)},
\ee
where we have again used Stirling's formula (\ref{eq:Stirling}). This result indicates 
that the energy denominators contribute a factor of
$n\hbar\omega/e=\Delta/e$ to $\alpha_{(b,n)}$, which for fixed $\Delta$ is
independent of $n$. Similar results are obtained for sequences involving individual
processes that lower the ``photon'' number in steps larger than one.\footnote{One example
is the case, where for an even value of $n$ we combine $n/2$ processes with matrix
elements $\propto \alpha^2$ that individually lower the photon number by two. In this
case the energy denominator can take the form $(n-2)!!(\hbar\omega)^{n/2-1}%=(n-2)(n-4)\cdots 1 (\hbar\omega)^{n/2-1}
= (n/2-1)! (2\hbar\omega)^{n/2-1}\simeq \sqrt{\pi (n-2)}[(n-2)\hbar\omega/e]^{n/2-1}$. It
contributes a factor of $\sqrt{\Delta/e}$ to $\alpha_{(b,n)}$, which is again independent
of $n$.}
Apart from the energy denominators also the matrix elements contribute to
$\alpha_{(b,n)}$. In the present example of a lattice with modulated lattice depth, we 
must expect that the $1/n$-dependence of the matrix elements (\ref{eq:V1}) and
(\ref{eq:V2}) leads to an increase of $\alpha_{(b,n)}$ with $n$. This effect is not
captured by the rotating-wave approximation, which takes these matrix elements into
account in linear order only. It can explain the behavior visible in
Fig.~\ref{fig:spectrum} that for lower driving 
frequencies larger driving strengths are required for significant resonant excitation.

We have started our perturbation expansion from the Hamiltonian $\Ho'(q,t)$ given by
Eq.~(\ref{eq:Hp}). In order to systematically improve the result (\ref{eq:M0}) obtained
within the rotating wave approximation, one can also start from the transformed 
Hamiltonian $\Ho''(q,t)$ given by Eq.~(\ref{eq:Hpp}). In this case we would recover the
result (\ref{eq:Crw}) already in first order. Note that the coupling matrix element
(\ref{eq:Crw}) contains infinite powers of the matrix element (\ref{eq:V0}), while it is
linear in the matrix elements (\ref{eq:V1}) and (\ref{eq:V2}). Transforming from 
$\Ho'(q,t)$ to $\Ho''(q,t)$, thus, corresponds to a resummation of part of the
perturbation series obtained for $\Ho'(q,t)$ to infinite order.

\section{Conclusions}
We investigated limitations of the low-frequency approximation that 
underlies typical protocols of Floquet engineering in systems of ultracold atomic quantum
gases in driven optical lattices. We stressed that already single-particle processes can
lead to unwanted transfer to excited orbital states beyond the low-frequency 
approximation. In order to illustrate this fact, we studied the example of a
one-dimensional optical lattice driven by a modulation of the lattice depth. For that
purpose we combined two different approaches. On the one hand, we computed excitation
spectra of the driven system by numerical means. On the other hand, we estimated the
effective coupling parameters for resonant interband transitions using an analytical
approach involving perturbation theory in Floquet space. The latter approach is able to
explain important features of the numerically computed spectra, like a momentum-dependent
suppression of transitions to the first excited band. 
The most important result is, however, the prediction of a threshold value of the driving
strength, below which interband excitations are suppressed exponentially for large $n$, 
that is for large inverse driving frequencies. 
 
We expect that this exponential suppression of interband heating with $n$ for
below-threshold driving is a rather general feature, which is found also for lattices 
driven by other forms of periodic forcing. Namely, the arguments that we employed to 
motivate the form (\ref{eq:scaling}) of the effective coupling matrix element $C_{b,n}$
for an $n$-``photon'' interband excitation process are rather general.
 
One example for another driving scheme is the shaken optical lattice investigated in
Ref.~\cite{WeinbergEtAl15}. For this system the driving strength $K$ can be defined as
the amplitude of the potential modulation between neighboring lattice sites in the 
comoving lattice frame, carrying the dimension of an energy. Using the above arguments 
together with the perturbation theory worked out in appendix of 
Ref.~\cite{WeinbergEtAl15}, the threshold driving strength can be evaluated to read 
\be
K_{(1,n)}=\frac{2\eta_{10}\Delta}{e},
\ee
where $\eta_{01}$ is a dimensionless matrix element describing the coupling between the two
lowest bands.\footnote{The suppression of transitions with even $n$ discussed in
Ref.~\cite{WeinbergEtAl15} is captured by the prefactor $B_{(1,n)}$, which obeys
$B_{(1,n)}\propto \eta_{10} K$ for odd $n$ and $B_{(1,n)}\propto 
\sin(aq)\eta_{10} K J/(\hbar\omega) $ for even $n$, where $J$ is a linear combination of 
the tunneling matrix elements of the lowest and the first excited band.} 
However, in the shaken optical lattice the physics of the system is mainly determined by
the scaled driving amplitude $K/(\hbar\omega)$. The threshold value for this relevant
quantity grows linearly with the ``photon'' number, 
$[K/(\hbar\omega)]_{(1,n)}=\Delta/(e\hbar\omega)=n/e$. 

We can now draw two conclusions concerning interband heating processes in Floquet driven 
optical lattices: On the one hand, heating processes to excited orbital states are a
relevant heating channel in periodically driven optical lattices already on the
single-particle level. On the other hand, such heating can be suppressed efficiently,
provided that both (i) the driving frequency is low enough, so that heating corresponds to
$n$-``photon'' transitions with large $n\gg1$, and (ii) the driving strength remains
below the threshold value. Our results contribute to a theoretical foundation of Floquet
engineering in periodically driven lattices. 

Relevant questions to be addressed in future work concern interband heating rates induced 
by driving schemes that were used to Floquet engineer artificial gauge fields. This 
includes driving functions involving higher harmonics of the driving frequency, as they
were used in the experiments described in Refs.~\cite{StruckEtAl12,StruckEtAl13}, as well 
as lattices that are driven by a moving running waves in order to create the Hofstadter 
Hamiltonian \cite{AidelsburgerEtAl15, AidelsburgerEtAl13, MiyakeEtAl13, KennedyEtAl15}. 
Furthermore, it will be crucial to understand in how far such single-particle heating 
channels are modified in systems of interacting particles. Also heating induced by two-particle 
scattering processes deserves further investigation.

\begin{acknowledgments}
This work was inspired by joint previous work with C. \"Olschl\"ager, K. Sengstock, 
S. Prelle, J. Simonet, M. Weinberg. C.S.\ acknowledges support from the Studienstiftung
des deutschen Volkes. 
\end{acknowledgments}

\bibliography{mybib}

\end{document}